\documentclass[
  aps,
  prb,
  reprint,
  superscriptaddress,
  amsmath,amssymb,
]{revtex4-2}
\usepackage{graphicx}
\usepackage{bm}
\usepackage{siunitx}
\usepackage{xcolor}
\usepackage{hyperref}
\usepackage{natbib}
\usepackage{braket}
\usepackage[normalem]{ulem}

\begin{document}

\title{A flux-controlled two-site Kitaev chain}

\author{Ivan Kulesh}
\altaffiliation{These authors contributed equally to this work.}
\affiliation{QuTech and Kavli Institute of Nanoscience, Delft University of Technology, Delft, 2600 GA, The Netherlands}
\author{Sebastiaan L. D. ten Haaf}
\altaffiliation{These authors contributed equally to this work.}
\affiliation{QuTech and Kavli Institute of Nanoscience, Delft University of Technology, Delft, 2600 GA, The Netherlands}

\author{Qingzhen Wang}
\affiliation{QuTech and Kavli Institute of Nanoscience, Delft University of Technology, Delft, 2600 GA, The Netherlands}

\author{Vincent P. M. Sietses}
\affiliation{QuTech and Kavli Institute of Nanoscience, Delft University of Technology, Delft, 2600 GA, The Netherlands}

\author{Yining Zhang}
\affiliation{QuTech and Kavli Institute of Nanoscience, Delft University of Technology, Delft, 2600 GA, The Netherlands}

\author{Sebastiaan R. Roelofs}
\affiliation{QuTech and Kavli Institute of Nanoscience, Delft University of Technology, Delft, 2600 GA, The Netherlands}

\author{Christian G. Prosko}
\affiliation{QuTech and Kavli Institute of Nanoscience, Delft University of Technology, Delft, 2600 GA, The Netherlands}

\author{Di Xiao}
\affiliation{Department of Physics and Astronomy, Purdue University, West Lafayette, Indiana 47907, USA}

\author{Candice Thomas}
\affiliation{Department of Physics and Astronomy, Purdue University, West Lafayette, Indiana 47907, USA}

\author{Michael J. Manfra}
\affiliation{Department of Physics and Astronomy, Purdue University, West Lafayette, Indiana 47907, USA}
\affiliation{School of Materials Engineering, Purdue University, West Lafayette, Indiana 47907, USA}
\affiliation{Elmore School of Electrical and Computer Engineering, Purdue University, West Lafayette, Indiana 47907, USA}

\author{Srijit Goswami}
\email{S.Goswami@tudelft.nl}
\affiliation{QuTech and Kavli Institute of Nanoscience, Delft University of Technology, Delft, 2600 GA, The Netherlands}

\date{\today}

\begin{abstract}
    In semiconducting-superconducting hybrid devices, Andreev bound states (ABSs) can mediate the coupling between quantum dots (QDs), allowing for the realisation of  artificial Kitaev chains. In order to engineer  Majorana bound states (MBSs) in these systems, one must control the energy of the ABSs. In this work, we show how extended ABSs in a flux tunable Josephson junction can be used to control the coupling between distant quantum dots separated by $\simeq~$\qty{1}{\micro \meter}. In particular, we demonstrate that the combination of electrostatic control and phase control over the ABSs significantly increases the parameter space in which MBSs are observed. Finally, by employing an additional spectroscopic probe in the hybrid region between the QDs, we gain information about the spatial distribution of the Majorana wave function in a two-site Kitaev chain. 
   
\end{abstract}

\maketitle

\section{Introduction}
\label{sec:ftpmm_introduction}

Quantum dots (QDs) with an induced superconducting (SC) coupling can be used to create solid-state quantum entanglers \cite{recher2001andreev, chtchelkatchev2002bell, samuelsson2003orbital, busz2017spin, brange2017minimal}, as well as to implement quantum gates for spin-qubits  \cite{leijnse2013coupling,spethmann2024high}.
Moreover, QDs with both a superconducting coupling and a hopping interactions offer a platform for constructing an artificial Kitaev chain \cite{kitaev2001unpaired, sau2012realizing}, hosting Majorana bound states (MBSs). 
The minimal chain is a system of two QDs, which can be tuned to host so-called poor man's Majoranas (PMMs) \cite{leijnse2012parity, tsintzis2022creating}. 
While not topologically protected from perturbations, these states are expected to exhibit Majorana properties, such as non-Abelian exchange statistics \cite{liu2023fusion,tsintzis2024majorana}. 
A crucial prerequisite to implement a PMM is the ability to control the coupling between spin-polarized QDs.
It was demonstrated that a proximitized semiconducting-superconducting hybrid region, hosting Andreev bound states (ABS), is an excellent mediator to couple the QDs \cite{liu2022tunable, bordin2023tunable, wang2023triplet, liu2024coupling}. 
The nature and magnitude of this coupling can be tuned by changing the electrostatic potential of the hybrid region, as well as by changing the orientation of the external magnetic field \cite{bordin2023tunable, wang2023triplet}.
Both were recently explored in experiments demonstrating the realization of the PMM states \cite{dvir2023realization, ten2023engineering}.
\newline\newline
Here, we further explore the role of ABSs in realizing PMM states by exploiting a novel geometry in a two-dimensional electron gas (2DEG) hybrid system. 
Motivated by recent theory work \cite{luna2024flux}, we utilize ABSs coupled to two SC electrodes embedded in a loop, such that the phase difference between the electrodes can be controlled with an applied perpendicular magnetic field.
Taking advantage of the long coherence length of the 2DEG, we are able to couple QDs separated by about $\qty{1}{\micro \meter}$.
We observe that the superconducting phase difference changes the effective coupling between two QDs and allows for finding PMM sweet spots within a continuous range of the ABS chemical potential.
This is an improvement on systems without this phase control, where sweet spots arise only at two discrete points of the ABSs chemical potential~\cite{liu2022tunable, bordin2023tunable}.
Despite the relatively large separation of the two QDs, an induced inter-dot coupling on the order of 10-$\qty{30}{\micro e\volt} $ is extracted from the spectroscopic measurements.
This demonstrates a clear advantage of using ABSs in proximitized semiconductors for providing a long-range coupling between QDs and relaxes spatial restrictions on PMM-device design.
Lastly, exploiting the flexible 2DEG architecture, we utilize a spectroscopic probe connected to the proximitized ABS segment. 
This additional probe allows us to study the spatial distribution of the PMM wavefunctions in the strongly coupled  regime~\cite{liu2023enhancing}.

\begin{figure}[ht!]
  \centering
  \includegraphics[width=\columnwidth]{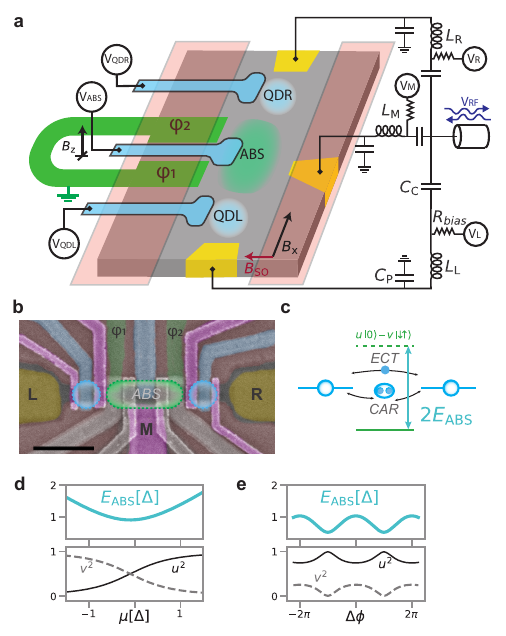}
  \caption{
  Schematic of the device measured (\textbf{a}) and false-colored SEM image (\textbf{b}): a grounded superconducting loop (green) is used to create a phase difference between two terminals $\phi_1$ and $\phi_2$, extending into the ABS region in a one-dimensional channel, formed by two depletion gates (red). 
  Plunger gates (blue) are used to control QDs and ABS chemical potential. 
  Three normal spectroscopic terminals (yellow) are connected to  resonators, formed by off-chip coil inductors $L_{\mathrm{X}}$ and the parasitic capacitances of bondwires $C_{\mathrm{P}}$, with coupling capacitors $C_{\mathrm{C}}$  allowing for a multiplexed read-out. 
  Cutter gates (pink, shown only in \textbf{b}) are used to confine quantum dots and define tunnel barriers. The scale bar in (\textbf{b}) is \qty{500}{\nano \meter}. 
  The experiment is shown schematically in (\textbf{c}): ABS (green) mediates ECT and CAR between quantum dots.
  The amplitudes of those non-local processes depend on the ABS energy $E_\mathrm{ABS}$ and its coherence factors $u$ and $v$, which, in turn, can be controlled by either the ABS chemical potential $\mu$ (\textbf{d}) or the phase difference $\Delta \phi$ (\textbf{e}). 
  }
  \label{fig1}
\end{figure}

\section{Device Design \& Characterization}
\label{sec:ftpmm_design}

\begin{figure}[ht!]
  \centering
  \includegraphics[width=\columnwidth]{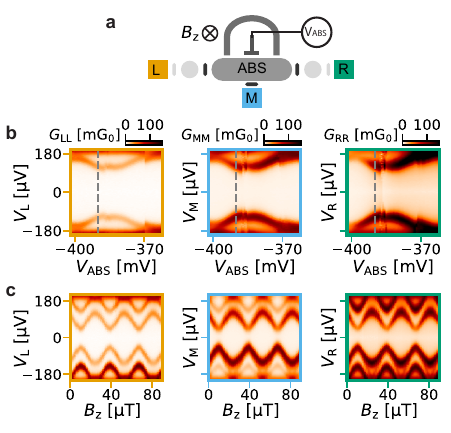}
  \caption{
  Schematic of the device with three tunneling spectroscopy terminals used to probe an extended ABS (\textbf{a}). The four outermost gates, used for forming the quantum dots, are not energized. 
  Top panel (\textbf{b}): tunneling spectroscopy measurements acquired while varying the ABS chemical potential.  
  Bottom panel (\textbf{c}): tunneling spectroscopy measurements, varying the out-of-plane field threading the SC loop, at the $V_{\mathrm{ABS}}$ value indicated by the dashed line in (\textbf{b}). For both panels we plot conductance $G_{\mathrm{XX}}=dV_{\mathrm{X}}/dI_{\mathrm{X}}$.
  The periodicity in field (\qty{28}{\micro \tesla}) agrees well with the loop area (\qty{60}{\micro \meter \squared}). 
  Both measurements are performed at zero in-plane magnetic field.
  }
  \label{fig2}
\end{figure}

The device is implemented on an InSbAs two-dimensional electron gas capped with $\qty{7}{\nano \meter}$ epitaxial aluminum \cite{moehle2021insbas}. 
Operating in depletion mode, the first layer of gates (red, Fig.\ref{fig1}a,b) define a one-dimensional channel connected to three spectroscopy terminals (yellow). 
Voltages are applied to the plunger gates (blue), with
$V_\mathrm{QDL}$ and $V_\mathrm{QDR}$ tuning the respective QDs' chemical potentials, while $V_\mathrm{ABS}$ controls the chemical potential of the hybrid region.
The plunger gates and the tunnel gates (purple) are situated in the second and third layers.\newline

Two superconducting terminals (green)  are connected in a loop and protrude into the channel, establishing an extended proximitized region.
The superconducting loop is kept grounded, so that an out-of-plane field $B_\mathrm{z}$ controls the phase difference $\Delta \phi = \phi_2 - \phi_1$ between the two SC electrodes. 
Measurements are performed using sub-GHz off-chip resonators, which are connected to the corresponding tunnel probes (see supplementary information \cite{suppl} for schematics).
The resonators can be probed simultaneously using a multiplexed reflectometry setup \cite{hornibrook2014frequency}.
The complex reflected RF signal of each resonator is converted to the single real value $\tilde{V}_{RF}^{\mathrm{X}}$ by performing a rotation in the complex plane \cite{suppl}. 
This signal is representative of the device conductance \cite{reilly2007fast, jung2012radio,razmadze2019radio}.
Each resonator is connected to a bias-tee $R_{\mathrm{bias}}$, allowing for applying a bias voltage to each spectroscopic normal lead, as well as for measuring the flow of current through each lead.
This additionally allows extracting the device conductance $G_{xx}=dV_{x}/dI_{x}$ using a standard low-frequency lock-in technique. 
The measurements are performed in a dilution refrigerator with a base temperature of \SI{20}{\milli \kelvin}.

The QDs are coupled by a hopping interaction through elastic co-tunneling (ECT) and by a pairing interaction arising from crossed Andreev reflection (CAR), schematically illustrated in Fig.\ref{fig1}c. 
The amplitudes of both these non-local processes depend on the ABS energy $E_\mathrm{ABS}$ \cite{liu2022tunable}. 
Additionally, ECT and CAR respond differently to the charge character of the ABS (determined by the coherence factors $u$ and $v$). 
Thus, when varying the ABS parameters with either the chemical potential $\mu$ or the phase $\Delta \phi$ (schematically illustrated in Fig.\ref{fig1}d,e), one expects to be able to control the ECT to CAR ratio \cite{luna2024flux}.

To mediate ECT and CAR between the distant QDs, an ABS is required that extends throughout the entire hybrid region in between them.
We first establish the presence of such an extended state in this device, particularly considering the relatively large length of this region -- \qty{700}{\nano \meter}, comparable to the SC coherence length in similar systems~\cite{mayer2019superconducting}. 
This is achieved by performing tunneling spectroscopy measurements from three sides (left, middle and right), while the four outermost barriers used for forming quantum dots are not energized (Fig.\ref{fig2}a). 
We separately vary either the ABS chemical potential by changing the voltage $V_{\mathrm{ABS}}$, or the phase $\Delta \phi$ by applying a magnetic field $B_\mathrm{z}$. 
We observe that the spectrum shows a correlated dependence from all three terminals, both as a function of gate voltage (Fig.\ref{fig2}b) and magnetic field (Fig.\ref{fig2}c), implying that a single quantum state is accessible to both quantum dots.

\section{Tuning into PMM regime}
\label{sec:ftpmm_pmm}

\begin{figure}[ht!]
  \centering
  \includegraphics[width=\columnwidth]{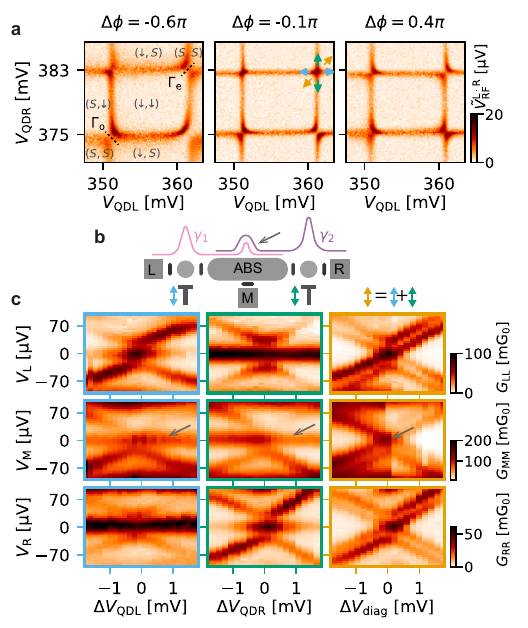}
  \caption{
  Tuning into the PMM sweet spot and verifying the PMM spectrum.
  Top panel (\textbf{a}): charge stability diagrams, shown in correlated voltage $\tilde{V}_{RF}^{\mathrm{L} \cdot \mathrm{R}} = \sqrt{ \tilde{V}_{RF}^{\mathrm{L}} \cdot \tilde{V}_{RF}^{\mathrm{R}}}$ (each component measured from the Coulomb blockade \cite{suppl}), varying the superconducting phase difference $\Delta \phi \left( B_\mathrm{z} \right)$. Combined QD states are indicated in brackets. States, coupled by $\Gamma_{\mathrm{o}}$ ($\Gamma_{\mathrm{e}}$) are connected with the dashed lines. (\textbf{b}): Schematic of the detuning experiment, showing PMM wave-functions $\gamma_1$ and $\gamma_2$, which can reside partially in the ABS region. (\textbf{c}): tunneling spectroscopy, measured from all three probes, plotted in conductance $G_{\mathrm{XX}}$. Note the signal at zero bias measured at the middle probe, highlighted with the arrows.}
  \label{fig3}
\end{figure}

Having established the presence of extended ABSs in our device, we proceed with forming the PMM system. 
First, we energize the additional outermost gates to define the QDs.
An in-plane magnetic field $B_{\mathrm{x}}=\qty{150}{\milli \tesla}$ is applied, in order to spin-polarize the QDs.
To achieve strong coupling between the QDs, the innermost gates are set to have relatively high tunneling between the QDs and the hybrid region~\cite{liu2023enhancing, zatelli2023robust}.
In this regime, the QDs are commonly described as Yu-Shiba-Rusinov (YSR) states \cite{meng2009self, grove2009superconductivity, deacon2010tunneling, lee2014spin, jellinggaard2016tuning}, as observed in spectroscopy measurements \cite{suppl}.
Depending on the dot chemical potential, the ground state of each QD is either a $\lvert \downarrow \rangle$ doublet or a singlet $\lvert S \rangle$ superposition of the empty and double-occupied QD state.
With this description of the QD states, the simple picture of ECT and CAR interactions can be extended.
Considering the combined state of the two QDs, two types of effective interactions can be defined~\cite{ten2023engineering,liu2023enhancing}: spin-conserving $\Gamma_{\mathrm{o}}$, coupling $\lvert S,\downarrow  \rangle$ with $\lvert \downarrow, S  \rangle$, and spin non-conserving $\Gamma_{\mathrm{e}}$ which couples $\lvert \downarrow,\downarrow  \rangle$ with $\lvert S,S  \rangle$, see Fig.\ref{fig3}a. 
Notably, those quantities can be expressed via ECT and CAR amplitudes $t$ and $\Delta$. 
Thus, $\Gamma_{\mathrm{o}}$ is a linear combination of spin-conserving terms $t_{\downarrow \downarrow}, t_{\uparrow \uparrow}, \Delta_{\uparrow \downarrow}, \Delta_{\downarrow \uparrow}$ (note that the CAR couples opposite spins), while $\Gamma_{\mathrm{e}}$ can be expressed via spin-flipping terms $t_{\uparrow \downarrow}, t_{\downarrow \uparrow}, \Delta_{\uparrow \uparrow}, \Delta_{\downarrow \downarrow}$.  
Without the spin-orbit interaction present, or when the external magnetic field is applied alongside the spin-orbit field $B_{\mathrm{SO}}$, only $\Gamma_{\mathrm{o}}$ is significant. 
Therefore, we apply the in-plane magnetic field alongside the dot-dot axis, perpendicular to the direction of the $B_{\mathrm{SO}}$ (see \cite{suppl}).~\newline

The different types of couplings between the QDs are revealed in charge stability diagrams (CSD), obtained by sweeping the QD plunger voltages and measuring reflected RF signals from the left and right normal leads, Fig.\ref{fig3}a. 
We record the CSDs while varying $B_\mathrm{z}$, which can be converted to the phase difference $\Delta \phi$ across SC electrodes (we assign $\Delta \phi = 0$ to the point of maximum $E_{\mathrm{ABS}}$ \cite{suppl}). 
Measured avoided crossings demonstrate the $\Gamma_{\mathrm{o}}$ ($\Gamma_{\mathrm{e}}$) coupling dominating, depending on whether the avoided crossing is (anti-)diagonal. 
In this example, we observe that varying $\Delta \phi$ indeed can change the coupling regime from $\Gamma_{\mathrm{e}} > \Gamma_{\mathrm{o}}$ to $\Gamma_{\mathrm{e}} < \Gamma_{\mathrm{o}}$ for the top-right transition, with the modulation being $2\pi$-periodic in  $\Delta \phi$~\cite{suppl}. 
This ability to change the coupling regime with the phase difference originates from the fact that $\Delta \phi$ affects the ABS coherence factors $u$ and $v$ (Fig.\ref{fig1}), contributing in a different manner to ECT and CAR \cite{liu2022tunable,luna2024flux}, and, consequently, to $\Gamma_{\mathrm{e}}$ and $\Gamma_{\mathrm{o}}$. 
Moreover, the presence of Zeeman field and the spin-orbit interaction in the proximitized region further affects the interplay between $\Delta \phi$ and the spin-split ABS spectrum \cite{yokoyama2014anomalous, van2017zeeman, tosi2019spin}, additionally affecting the spin-conserving(-flipping) ECT $t_{\sigma_1 \sigma_2}$ and CAR $\Delta_{\sigma_1 \sigma_2}$ amplitudes \cite{bordin2023tunable}.\newline

The point where $\Gamma_{\mathrm{e}} = \Gamma_{\mathrm{o}}$ corresponds to the so-called PMM sweet spot \cite{liu2023enhancing, ten2023engineering}, with the dot transitions crossing in straight lines.
We proceed to verify the sweet spot conditions by performing the spectroscopy measurements (Fig.\ref{fig3}b,c) while detuning either one or both QDs.
As expected \cite{leijnse2012parity}, the zero-bias conductance peak persists when detuning only one QD and splits from zero energy when detuning both QDs.
The minimum energy of the excited states allows for estimating the coupling amplitudes to be on the order of $\qty{18}{\micro\volt}$ (and up to $\qty{30}{\micro\volt}$ for the regime described in SI).
This is comparable to previously reported values \cite{ten2023engineering,zatelli2023robust}, indicating that the increased length of the ABS segment does not significantly impact the interaction strength.
Operating in the strong coupling regime, a finite overlap of the PMM wave functions inside the ABS segment is expected~\cite{liu2023enhancing, luna2024flux} but has not yet been directly probed.
Leveraging the flexibility of the 2DEG platform, we utilize the spectroscopic probe of the ABS region to study this.  
As shown in Fig.\ref{fig3}c, we find that a zero-bias conductance peak is also clearly visible in the conductance $G_{\mathrm{MM}}$.
This is an indication that the PMM wave functions $\gamma_1$ and $\gamma_2$ both reside partially in the ABS region, such that electron transfer from the lead M to the delocalized zero-energy fermionic mode is possible.

It is important to note that the observed wave function overlap in the ABS region is not expected to be detrimental for the device performance. 
In contrast, the outer QDs must support only a single PMM wave function to ensure optimal conditions to explore the MBS physics\cite{liu2023enhancing,luna2024flux,tsintzis2024majorana}.

\section{Exploring the gate-phase parameter space}
\label{sec:ftpmm_gate}
The section above demonstrates that the relative amplitudes of $\Gamma_{\mathrm{e}}$ and $\Gamma_{\mathrm{o}}$ couplings can be accurately controlled through controlling the superconducting phase difference.
Now, we proceed to explore how this can be used to complement the previously established control utilising the ABS chemical potential~\cite{dvir2023realization,bordin2023tunable,wang2023triplet}.
As with phase, this tunability is achieved through the dependence of the ABS energy $E_{\mathrm{ABS}}$, as well as the coherence factors $u$ and $v$, on ABS gate voltage \cite{liu2022tunable}. 
Moreover, one expects that the dependence of the ABS parameters on the phase difference $\Delta \phi$ is modified when varying the ABS chemical potential $\mu$ \cite{luna2024flux}. 
Qualitatively, this can be pictured as the junction transparency being a function of $\mu$. 
This further motivates us to explore the two-dimensional $V_{\mathrm{ABS}}$,~$\Delta \phi$ parameter space.
\begin{figure}[ht!]
  \centering
  \includegraphics[width=\columnwidth]{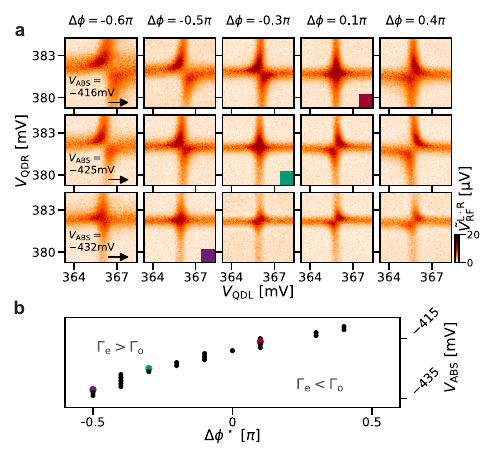}
  \caption{
  The sweet spot location in the ABS gate - SC phase difference space. We capture charge stability diagrams (\textbf{a}) for a set of fixed $V_\mathrm{ABS}$ voltages while varying $\Delta \phi$. The sweet spot $\Delta \phi^*$ is then determined for each $V_\mathrm{ABS}$ as a crossover between $\Gamma_{\mathrm{o}}$ and $\Gamma_{\mathrm{e}}$, corresponding to the straight QD charge transition lines, and marked with a colored square. We repeat this procedure for multiple $V_\mathrm{ABS}$ values and plot the extracted $\Delta \phi^*$ in (\textbf{b}).
  }
  \label{fig4}
\end{figure}
The results of this experiment are summarized in Fig.\ref{fig4}. 
We record CSDs, focusing on a specific transition (corresponding to the top-right on the Fig.\ref{fig3}a), while varying $\Delta \phi \left(B_{\mathrm{z}} \right)$.
Comparing these CSDs for different $V_{\mathrm{ABS}}$, we observe that a PMM sweet spot, corresponding to the crossover between dominant $\Gamma_{\mathrm{e}}$ to $\Gamma_{\mathrm{o}}$, can be obtained for each $V_{\mathrm{ABS}}$ set-point in this range.
Moreover, small changes in the $V_{\mathrm{ABS}}$ result in a small change of the sweet spot phase difference  $\Delta \phi^*$. 
This demonstrates that for certain $\Delta \phi$,$V_{\mathrm{ABS}}$ range, the PMM sweet spot spans a continuous line in a two-dimensional space instead of being a single point at a specific gate (phase) value.
It is predicted that the sweet spot characteristics, such as the excitation gap, change systematically alongside $\Gamma_{\mathrm{e}} = \Gamma_{\mathrm{o}}$ line in the $\Delta \phi, V_{\mathrm{ABS}}$ plane. 
However, we have not observed any such systematic behavior, exploring multiple regimes throughout several device cool-downs (see SI). 
This discrepancy with the theoretical predictions \cite{luna2024flux} can be attributed to the presence of multiple states in the proximitized region (Fig.\ref{fig2}c), as well as the spin-splitting of said states.

\section{Conclusions}
\label{sec:ftpmm_gate}
We have demonstrated the possibility of strongly coupling two QDs via an extended Andreev bound state in a proximitized InSbAs 2DEG on a \qty{700}{\nano \meter}-long hybrid segment.
Obtained coupling amplitudes are on the same order of magnitude as reported in preceding studies~\cite{dvir2023realization, ten2023engineering} with short (150-\qty{200}{\nano \meter}) hybrid segments, thus relaxing constraints for future designs.
We found that embedding the hybrid region in a SC loop allows for a novel control knob over the effective ECT and CAR couplings between the two QDs with the superconducting phase difference.
Our design can be of interest for realizing a long-range tunable superconducting coupling for spin-qubit architectures~\cite{leijnse2013coupling}.
Here, we explored the phase control to tune the system to a set of sweet spots hosting PMM states.
Combined with control through the ABS chemical potential established in previous works, we found that sweet spots can be obtained along a continuous path in the gate-phase space.
Future studies can benefit from dedicated flux lines to control the interaction between QDs, which may reduce gate cross-coupling and enhance charge stability when compared with the electrostatic gate control.
Lastly, we studied the PMM wave functions in the ABS segment and observed their presence in tunneling spectroscopy, suggesting the states are not fully localized on the QDs.
These results expand the device geometry and support the understanding of coupled quantum dots in a PMM system.

\section*{Data Availability}

Raw data and scripts for plotting the figures in this publication are available from Zenodo \cite{zenodo}.
    
\begin{acknowledgements}
    The authors would like to thank O.W.B. Benningshof and J.D. Mensingh for technical assistance with the cryogenic electronics. The authors are grateful to Juan Daniel Torres Luna and Chun-Xiao Liu for helpful discussions. The research was
supported by the Dutch National Science Foundation
(NWO), Microsoft Corporation Station Q and a grant
from Top consortium for Knowledge and Innovation program (TKI). S.G. acknowledges financial support from the Horizon Europe Framework Program of the
European Commission through the European Innovation
Council Pathfinder grant no. 101115315 (QuKiT).
\end{acknowledgements}

\section*{Author Contributions}
The device was fabricated by Q.W. using a 2DEG heterostructure provided by D.X., C.T., and M.J.M., while S.L.D.t.H. and I.K. contributed to the device design and fabrication flow optimization. The experiment was devised by I.K. Initial device characterization was performed by Y.Z.
Experimental data was obtained by V.P.M.S, supervised by I.K. and S.L.D.t.H.
I.K., S.L.D.t.H., V.P.M.S., Y.Z., S.R.R., and C.G.P. designed the measurement setup. I.K. and S.L.D.t.H. wrote the manuscript with input from all authors. S.G. supervised the project.




\bibliography{ftpmm_bib}

\end{document}


\title{A flux-controlled two-site Kitaev chain}

\author{Ivan Kulesh}
\altaffiliation{These authors contributed equally to this work.}
\affiliation{QuTech and Kavli Institute of Nanoscience, Delft University of Technology, Delft, 2600 GA, The Netherlands}
\author{Sebastiaan L. D. ten Haaf}
\altaffiliation{These authors contributed equally to this work.}
\affiliation{QuTech and Kavli Institute of Nanoscience, Delft University of Technology, Delft, 2600 GA, The Netherlands}

\author{Qingzhen Wang}
\affiliation{QuTech and Kavli Institute of Nanoscience, Delft University of Technology, Delft, 2600 GA, The Netherlands}

\author{Vincent P. M. Sietses}
\affiliation{QuTech and Kavli Institute of Nanoscience, Delft University of Technology, Delft, 2600 GA, The Netherlands}

\author{Yining Zhang}
\affiliation{QuTech and Kavli Institute of Nanoscience, Delft University of Technology, Delft, 2600 GA, The Netherlands}

\author{Sebastiaan R. Roelofs}
\affiliation{QuTech and Kavli Institute of Nanoscience, Delft University of Technology, Delft, 2600 GA, The Netherlands}

\author{Christian G. Prosko}
\affiliation{QuTech and Kavli Institute of Nanoscience, Delft University of Technology, Delft, 2600 GA, The Netherlands}

\author{Di Xiao}
\affiliation{Department of Physics and Astronomy, Purdue University, West Lafayette, Indiana 47907, USA}

\author{Candice Thomas}
\affiliation{Department of Physics and Astronomy, Purdue University, West Lafayette, Indiana 47907, USA}

\author{Michael J. Manfra}
\affiliation{Department of Physics and Astronomy, Purdue University, West Lafayette, Indiana 47907, USA}
\affiliation{School of Materials Engineering, Purdue University, West Lafayette, Indiana 47907, USA}
\affiliation{Elmore School of Electrical and Computer Engineering, Purdue University, West Lafayette, Indiana 47907, USA}

\author{Srijit Goswami}
\email{S.Goswami@tudelft.nl}
\affiliation{QuTech and Kavli Institute of Nanoscience, Delft University of Technology, Delft, 2600 GA, The Netherlands}

\maketitle

\begin{widetext}
\tableofcontents
\clearpage
\section{Methods}

\subsection*{Device fabrication}

All devices were fabricated using techniques described in~\cite{moehle2021insbas}.
An aluminum ring with two narrow extended strips is defined in an InSbAs-Al chip by wet etching, followed by the deposition of two normal Ti/Pd contacts. 
After placing \SI{20}{nm} AlOx via atomic layer deposition (ALD), three Ti/Pd depletion gates are evaporated. 
Following a second ALD layer (\SI{20}{nm} AlOx), multiple Ti/Pd finger gates are evaporated. In a similar fashion, we define a third layer of gates. Finger gates in the second and third layers are used to define the barriers and tune the chemical potentials. 
The depletion gates define a quasi-1D channel with a width of about \SI{150}{nm}, contacted on each side and in the middle by a normal lead. The aluminum ring extends into the channel and induces superconductivity in the ABS section of the device, with an induced gap on the order of $\SI{200}{\upmu eV}$.
ABSs are found to be present over a large range of $V_{\mathrm{ABS}}$, the voltage applied to the gate covering the hybrid region.
Finger gates define QDs with charging energies around \SI{1}{mV}.

A single device was used to obtain the data presented in the main text.  Measurements were performed in a dilution refrigerator with a base temperature of \SI{20}{mK}. 
The main text contains information from a single cool-down, while the supplementary contains additional datasets from two preceding cool-downs (\Cref{fig_SI_fig3_ds2,fig_SI_fig4_ds2,fig_SI_fig4_ds1}).

 \subsection*{Transport measurements and data processing}

Measurements are performed using sub-GHz off-chip resonators connected to corresponding tunnel probes. 
Each resonator circuit is formed by a spiral superconducting inductor and the parasitic capacitances of the bond wires.
The resonators can be probed simultaneously using a multiplexed reflectometry setup\cite{hornibrook2014frequency}, with frequencies $f_{\mathrm{L, M, R}} = 723, 505, 248$~\unit{\mega \hertz}, determined by the variation in the inductances $L_{\mathrm{L, M, R}} = 0.2, 0.5, 1.5$~\unit{\micro \henry}. 
We convert the complex reflected RF signal of each resonator to the single real value $\tilde{V}_{RF}^{\mathrm{X}}$ by performing a rotation in the complex plane, see \Cref{fig_SI_RF}. 
The signals are expected to scale linearly with the device conductance, such that it expresses the same features\cite{reilly2007fast, jung2012radio,razmadze2019radio}.
To fully benefit from the increased acquisition speed, we use a rastering method (\Cref{fig_SI_RF_setup}), with an arbitrary waveform generator (AWG) applying a two-channel sawtooth signal to the dot plunger gates. 
The first channel has a single wave period, while the second channel has as many periods $N_2$ as the number of rows in the resulting charge stability diagram. 
The frequency $f_B$ of the fast channel is limited by the RC filter cut-off frequency. 
The number of points in each row $N_1$ is limited by the acquisition device memory $N_1 \cdot N_2$ and the acquisition bandwidth $f_B \cdot N_1 < BW$.
Sawtooth signals from the AWG are connected to the optical isolator, breaking any potential ground loops, and are combined with the constant DC offsets using a voltage divider. 
They are then fed into the device through a set of filtered low-frequency fridge lines.

On the AWG trigger event, the VNA records $N_1 \cdot N_2$ points such that the total acquisition time equals the period $T_A$ of the slow channel wave. 
The VNA measures the complex reflected signal amplitude, with the RF circuit designed so that the incident and reflected signals from the resonator are separated into two lines with the help of a directional coupler. 
Finally, the one-dimensional output data string from the VNA is mapped into a two-dimensional matrix, corresponding to the charge stability diagram.

Each resonator is connected to a bias-tee $R_{bias}$, thus allowing the application of a bias voltage to each spectroscopic normal lead, as well as measuring the current flowing through it. 
This allows us to extract the device conductances $G_{xx}$ using a standard low-frequency lock-in technique. 
We assume total line resistances to be on the order of \qty{10}{\kilo \ohm}, and neglect the voltage divider effect, which for currents $ I_{\mathrm{x}} < \qty{1}{\nano \ampere} $ results in $< \qty{10}{\micro \volt}$ voltage offsets. 

Instrumental offsets of the applied voltage biases are corrected by independently calibrating the spectroscopic measurements on each side. When applying a DC voltage to the specific tunnel probe, the other probes are kept at the offset voltage value.
Magnetic fields were applied using a 3D vector magnet.
Due to device instabilities or charge jumps, electrostatics of the QDs experience small drifts over the course of the measurements. 
Investigated orbitals were tracked while collecting the presented datasets.
Such drifts are the cause of small discrepancies in gate voltages between figures from the same dataset.

\setcounter{figure}{0}
\renewcommand{\thefigure}{S\arabic{figure}}
\newpage
\section{Supplementary Figure S1 to S10}

\begin{figure*}[ht!]
  \centering
  \includegraphics{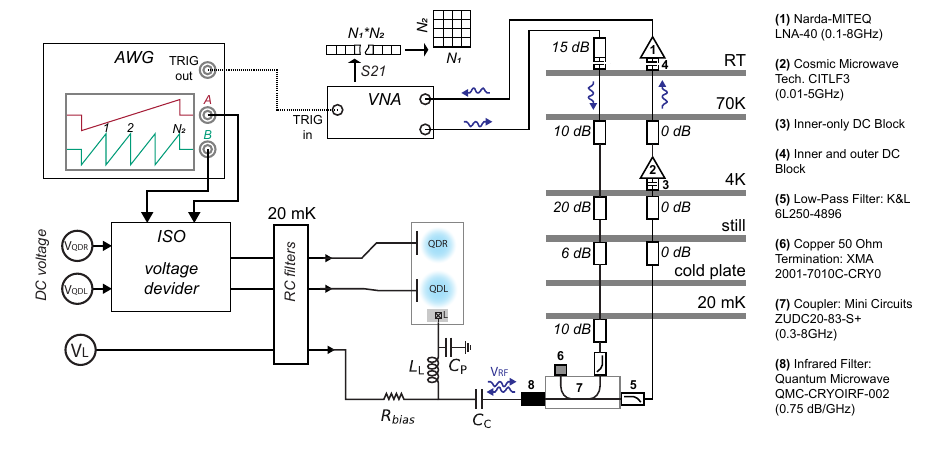}
  \caption{
  RF reflectometry and rastering setup. An arbitrary waveform generator (AWG) is used to provide two sawtooth signals, while the synchronized vector network analyzer (VNA) acquires a complex reflection coefficient. The microwave circuit (right side) separates incident and reflected waves, which is required to equilibrate effective noise temperature with attenuators and amplify the reflected signal. Only one lead resonator is shown for clarity. The device is represented schematically, showing only the QD plunger gates and the left normal lead.
  }
  \label{fig_SI_RF_setup}
\end{figure*}

\begin{figure*}[ht!]
  \centering
  \includegraphics{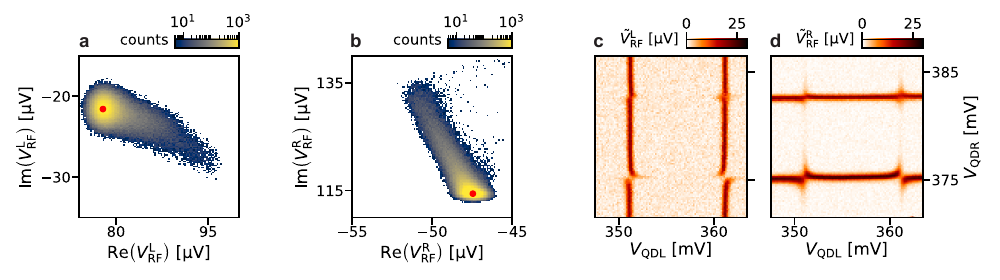}
  \caption{
  Signal extraction using RF reflectometry. We record the complex reflected voltage amplitude $V^X_\mathrm{RF}$ for each resonator $X$. Varying the plunger voltages in a sufficiently large range, we plot the complex signal on a two-dimensional histogram (\textbf{a},\textbf{b}). The maximum signal count (marked with the red dot) corresponds to the Coulomb blockade, the most prevalent state for the selected gate voltage range. We then extract a single value $\tilde{V}^X_\mathrm{RF}$ (\textbf{c},\textbf{d}), defined as the distance from a Coulomb blockade for each measured point.
  }
  \label{fig_SI_RF}
\end{figure*}

\begin{figure*}[ht!]
  \centering
  \includegraphics{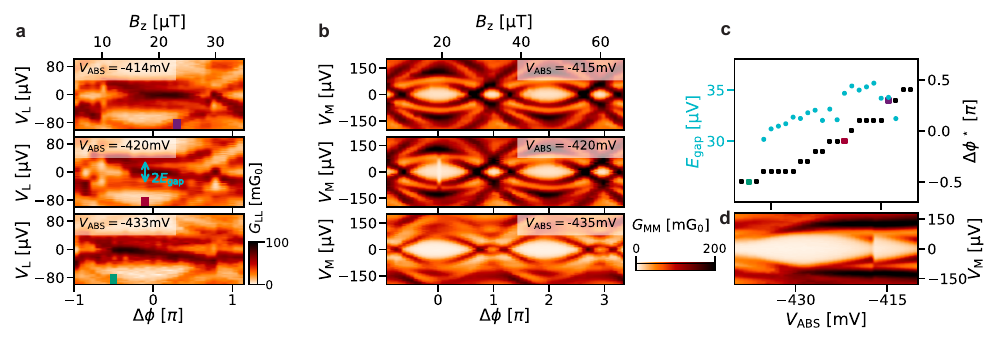}
  \caption{
  Tunneling spectroscopy of the PMM system around a sweet spot (\textbf{a}), measured from the left terminal. When detuning the phase difference $\Delta \phi$, we observe the splitting of the zero-bias peak, as the sweet spot exists only for the specific value $\Delta \phi^*$. For each $\Delta \phi$ point, quantum dot chemical potentials are adjusted to stay at the charge degeneracy point. Changing the ABS chemical potential with the $V_\mathrm{ABS}$ gate shifts the sweet spot position, also shown in (\textbf{c}) for a large set of $V_\mathrm{ABS}$ voltages, alongside the excitation gap $E_\mathrm{gap}$. To gain further insight into the relation between the sweet spot location and the system parameters, we record the ABS spectrum with the quantum dots in the Coulomb blockade as a function of the phase difference (\textbf{b}) and the gate voltage (\textbf{d}).
  }
  \label{fig_SI_det_field}
\end{figure*}

\begin{figure*}[ht!]
  \centering
  \includegraphics{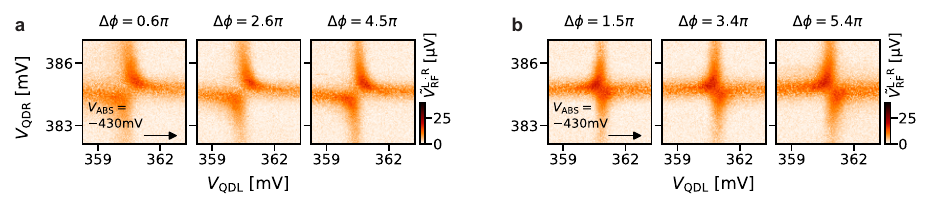}
  \caption{
  As the phase modulation of the ABS parameters is $2\pi$-periodic, we expect the same periodicity for the quantum dots interaction type and strength. To verify this, we record charge stability diagrams in a large $\Delta \phi$ range for two values of the ABS chemical potential $V_\mathrm{ABS}$ (\textbf{a}, \textbf{b}), confirming the expected behavior.
  }
  \label{fig_SI_periodicity}
\end{figure*}

\begin{figure*}[ht!]
  \centering
  \includegraphics{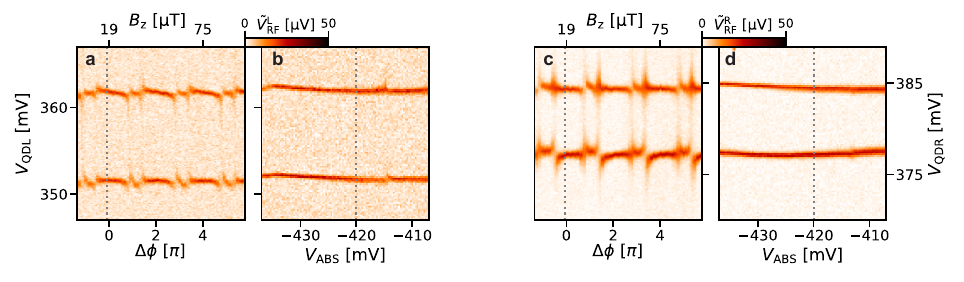}
  \caption{
  ABS not only acts as an interaction mediator between two quantum dots but also couples to an individual QD, modifying energy levels. ABS-QD coupling is revealed in the Coulomb peak shift as a function of the phase difference (\textbf{a} for QDL and \textbf{c} for QDR) and the ABS gate (\textbf{b},\textbf{d} for QDL and QDR respectively). Avoided crossings are present when the ABS energy approaches zero and are clearly visible when sweeping $\Delta \phi$.
  }
  \label{fig_SI_dot_ABS}
\end{figure*}

\begin{figure*}[ht!]
  \centering
  \includegraphics{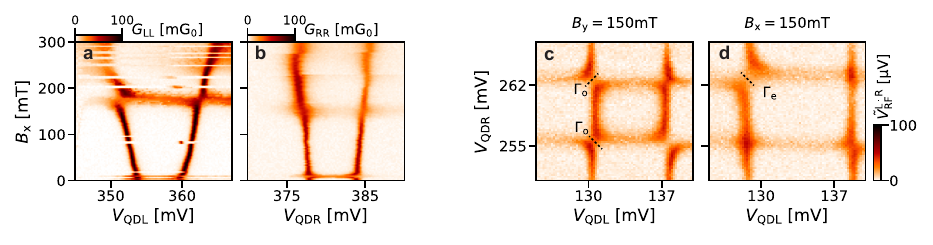}
  \caption{
  Coulomb peak evolution with the in-plane magnetic field (\textbf{a}, \textbf{b}) shows a clear Zeeman splitting, necessary for the quantum dot levels spin polarization. Depending on the in-plane field angle with respect to the spin-orbit field, charge stability diagrams demonstrate either only spin-conserving interaction $\Gamma_{\mathrm{o}}$ is allowed (\textbf{c}, $B_{\mathrm{SO}} \parallel B_Y$), or, additionally, spin non-conserving $\Gamma_{\mathrm{e}}$ (\textbf{d}, $B_{\mathrm{SO}} \perp B_X$).
  }
  \label{fig_SI_SOI}
\end{figure*}

\begin{figure*}[ht!]
  \centering
  \includegraphics{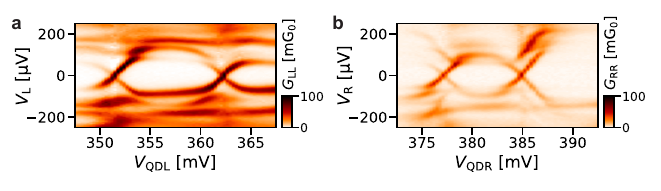}
  \caption{
  Tunneling spectroscopy of YSR states in the left quantum dot (\textbf{a}) demonstrates a strong coupling to the ABS. Thusly, the QDL ground state as either a singlet $\lvert S \rangle = u_{\mathrm{L}}\lvert 0 \rangle -v_{\mathrm{L}}\lvert \downarrow\uparrow\rangle$, or a doublet $\lvert \downarrow \rangle$. Tunneling spectroscopy of the right QD (\textbf{b}) shows a similar behavior.
  }
  \label{fig_SI_YSR}
\end{figure*}

\begin{figure*}[ht!]
  \centering
  \includegraphics{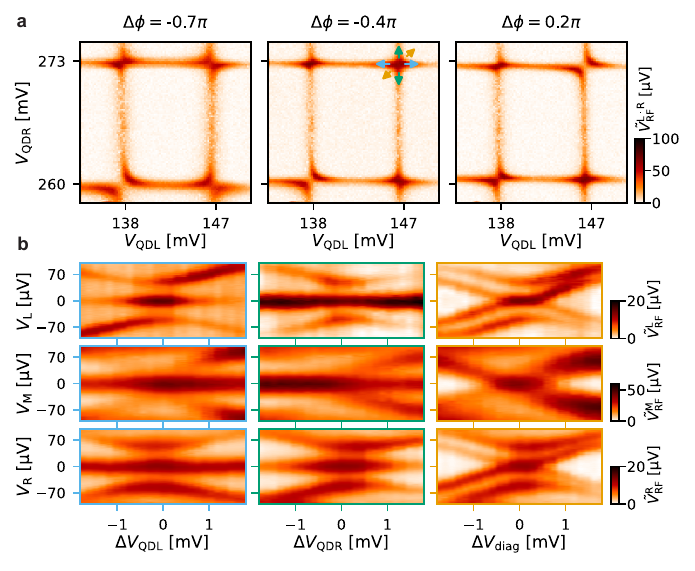}
  \caption{
  Additional data for device cool-down II, demonstrating tuning into a sweet spot with magnetic flux, shown with the charge stability diagrams while varying $\Delta \phi$ (\textbf{a}). We verify the PMM sweet spot conditions with the tunneling spectroscopy measurements from all three probes while detuning quantum dot plunger gates (\textbf{b}).
  }
  \label{fig_SI_fig3_ds2}
\end{figure*}

\begin{figure*}[ht!]
  \centering
  \includegraphics{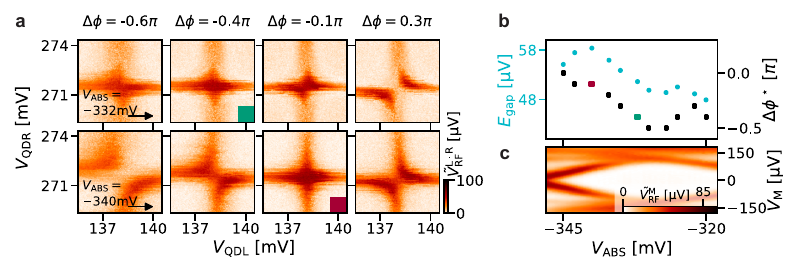}
  \caption{
  Sweet spot location in the $\Delta \phi$ - $V_{\mathrm{ABS}}$ parameter space for device cool-down II. The sweet spot phase difference $\Delta \phi^*$ is found by inspecting charge stability diagrams (\textbf{a}) while varying $\Delta \phi$ for a fixed $V_{\mathrm{ABS}}$ gate value and corresponds to the transition between dominant $\Gamma_{\mathrm{o}}$ and $\Gamma_{\mathrm{e}}$ (marked with a colored square). The same protocol is repeated for a range of $V_{\mathrm{ABS}}$ values with the extracted $\Delta \phi^*$, as well as the excitation gap in the sweet spot $E_\mathrm{gap}$, shown in (\textbf{b}). Corresponding ABS spectrum at $\Delta \phi=0$ as a function of $V_{\mathrm{ABS}}$ measured with the quantum dots in the Coulomb blockade is shown in (\textbf{c}).
  }
  \label{fig_SI_fig4_ds2}
\end{figure*}

\begin{figure*}[ht!]
  \centering
  \includegraphics{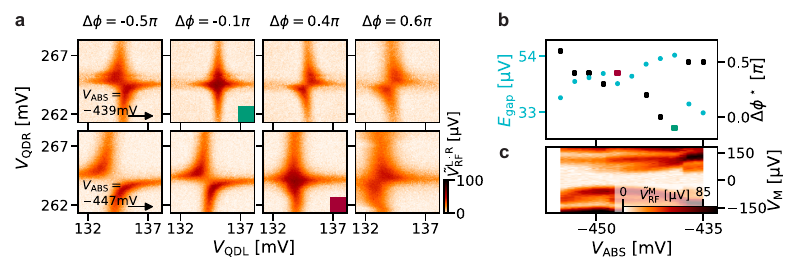}
  \caption{
  Sweet spot location in the $\Delta \phi$ - $V_{\mathrm{ABS}}$ parameter space for device cool-down I. The sweet spot phase difference $\Delta \phi^*$ is found by inspecting charge stability diagrams (\textbf{a}) while varying $\Delta \phi$ for a fixed $V_{\mathrm{ABS}}$ gate value and corresponds to the transition between dominant $\Gamma_{\mathrm{o}}$ and $\Gamma_{\mathrm{e}}$ (marked with a colored square). The same protocol is repeated for a range of $V_{\mathrm{ABS}}$ values with the extracted $\Delta \phi^*$, as well as the excitation gap in the sweet spot $E_\mathrm{gap}$, shown in (\textbf{b}). Corresponding ABS spectrum at $\Delta \phi=0$ as a function of $V_{\mathrm{ABS}}$ measured with the quantum dots in the Coulomb blockade is shown in (\textbf{c}).
  }
  \label{fig_SI_fig4_ds1}
\end{figure*}

\clearpage

\bibliography{ftpmm_bib} 

\end{widetext}
